\documentclass[aps,pre,superscriptaddress,onecolumn,showpacs,floatfix]{revtex4}
\usepackage{epsfig,graphicx,latexsym,amsmath,hyperref,amssymb}

\newcommand{\zz}{{\bf 0}}
\newcommand{\un}{{\bf 1}}

\newcommand{\xx}{{\boldsymbol{x}}}
\newcommand{\yy}{{\boldsymbol{y}}}
\newcommand{\z}{{\boldsymbol{z}}}
\newcommand{\pp}{{\boldsymbol{p}}}
\newcommand{\kk}{{\boldsymbol{k}}}
\newcommand{\qq}{{\boldsymbol{q}}}
\newcommand{\uu}{{\boldsymbol{u}}}

\newcommand{\FF}{{\boldsymbol{F}}}
\newcommand{\GG}{{\boldsymbol{G}}}

\newcommand{\XXi}{{\boldsymbol{\Xi}}}
\newcommand{\eeta}{{\boldsymbol{\eta}}}

\begin{document}

\title{A perturbative path integral study of active and passive tracer diffusion in fluctuating fields}

\author{Vincent D\'emery}
\affiliation{Laboratoire de Physique Th\'eorique, IRSAMC, Universit\'e Paul Sabatier, 118 Route de Narbonne, 31062 Toulouse Cedex 4, France}

\author{David S. Dean}
\affiliation{Laboratoire de Physique Th\'eorique, IRSAMC, Universit\'e Paul Sabatier, 118 Route de Narbonne, 31062 Toulouse Cedex 4, France}


\begin{abstract}
We study the effective diffusion constant of a Brownian particle linearly coupled to a thermally fluctuating  scalar field. We use a path integral method to compute the effective diffusion coefficient perturbatively  to
lowest order in  the coupling constant. This method can be applied to cases where the field is affected by the particle (an active tracer), and cases where the tracer is passive. Our results are applicable to a wide range of physical problems,  from a protein diffusing in a membrane to the dispersion of a passive tracer in a random potential. In the case of passive diffusion in a scalar field, we show that the coupling to the field can, in some cases, speed up the diffusion corresponding to a form of  stochastic resonance. Our results on passive diffusion are also confirmed via a perturbative calculation of the probability density function of the particle in a Fokker-Planck formulation of the problem. Numerical simulations on simplified systems corroborate our results.
\end{abstract} 

\maketitle

\section{Introduction}

Diffusion  in a quenched random medium is a problem which has been extensively investigated \cite{bou1990, dean2007}. An important physical problem is to understand how the long-time transport properties of a Brownian particle, notably its effective diffusion constant, are modified with respect to those of a homogeneous medium. Two types of problem with quenched disorder have been extensively studied. The first is  diffusion  in a medium where the local diffusion constant depends on spatial position and is taken to have some statistical distribution and the second is for a particle which is advected by a quenched random velocity field. The case where the random velocity field is derived from a potential is of importance as a toy model for spin glasses and glasses as it can exhibit a static spin glass like
transition \cite{mez1991, eng1993} and a dynamical structural glass-like transition
\cite{kin1993, fran1994, cug1996}. Variants of the toy model of diffusion in a random potential can exhibit a transition in their transport properties notably   
diffusion which is normal in the high temperature phase and becomes subdiffusive in the low temperature  phase \cite{bou1990, dean1998, touy2007, dean2008}. The onset of the subdiffusive regime is signaled by the vanishing of the late time diffusion constant. It should also be mentioned that above class of problems can also be related the problem of evaluating the macroscopic/effective electrical properties, such as conductivities or dielectric constants, of random conductors or dielectrics \cite{dean2007}. 

The case of diffusion in  dynamically evolving random media has perhaps received less attention. The most  widely studied problem of a time dependent nature is for a particle diffusing in a turbulent flow \cite{shr2000, fal2001}, and the particle here is a passive tracer and has no effect
on the flow field. Another example is for  a protein diffusing in a membrane, where the protein is subject
a force generated by membrane curvature or composition, and these quantities themselves fluctuate.
However in this case  the membrane  is in general  affected by the protein \cite{naji2009}. Naively one might expect that protein diffusion is speeded up by coupling to height or composition fluctuations in 
a membrane, however this is not the case, the feedback of the protein on the membrane configuration
actually slows down the protein diffusion with respect to a non-fluctuating homogeneous one. A general
question that arises in these sorts of problems is:  when does the fluctuating field speed up the diffusion 
of a tracer and when does it slow it down ? An important question, which has received much recent attention, is how the diffusion constant of a protein depends on its size. The classic 
hydrodynamic computation of Saffmann and Delbr\"uck \cite{saff1975} treats the protein as a solid cylinder in an incompressible layer of fluid (the lipid bilayer) sandwiched between another fluid (the water). In this formalism the protein diffusivity shows a weak logarithmic dependence on the cylinder radius. However the validity of this result has been called into question experimentally where a stronger dependence is reported \cite{gam2006}. A number of theoretical studies have suggested that \zz
a protein's diffusion could be modified by coupling to local membrane properties, such as composition
and curvature, that are not taken into account in a purely hydrodynamic model for a membrane
\cite{rei2005, naji2007, lei2008, naji2009, lei2010, dem2010a, dem2010b, dean2011}. One should 
bare in mind that protein coupling to local geometry and composition has also been postulated as
a mechanism for inter-protein interactions in biological membranes \cite{gou1993, sac1995}.

In this paper we will consider a particle, whose position is denoted by $\xx(t)$, diffusing in a scalar field $h\psi$, where $h$ is  a parameter controlling the coupling of the field to the particle. Thus the particle drift is given by $\uu= -\kappa_{\xx}h\nabla\psi$, where $\kappa_{\xx}$ denotes the bare diffusion
constant of the particle. The dynamics of both the particle and the field are overdamped and stochastic. Our aim is to evaluate how the effective diffusion coefficient for the particle is modified by its coupling to the field. A path integral approach allows us to perform a perturbative computation if the coupling constant between the particle and the field is small for a wide range of physically different problems. 
Within this formalism we recover some previous results on the diffusion of active tracers where the 
whole system (particle plus field) obeys detailed balance \cite{dean2011}. However the method also
allows us to analyze passive diffusion and a continuum of intermediate models with varying degrees 
of feedback of the particle on the field and for thermal fluctuations on the particle and field that
are not necessarily at the same temperature. This intermediate range of models could apply to
the study of tracers in non-thermally driven fields, and have applications to active colloidal systems \cite{pal2010}, swimmers \cite{loi2008} or systems where the tracer is heated with an external heat source such as a laser \cite{jol2011}. The basic method may also be useful to study systems where the
field itself is non-thermally driven, for instance lipid-bilayers where the field is driven an electric field
\cite{lac2009}.

The formalism also allows us to see how the 
relative time scales between the tracer and fluctuating field affect the effective diffusion constant of the tracer.

Our results for the effective diffusion constant show very different effects  for active and passive diffusion, and diffusion in a quenched potential.  In pure active diffusion, where the stochastic equations of motion obey detailed balance, the diffusion is always slowed down. For passive diffusion, if the field evolution is slow, the particle slows down, as if was evolving in a quenched potential \cite{dean2007}. As the field evolution is speeded up the diffusion constant increase and 
then reaches a maximum, in a manner reminiscent of stochastic resonance \cite{ben1981, hang1998, well2004, alc2004}. As the field evolution rate increases still further, the diffusion constant diminishes and eventually reaches the bare value it would have in the absence of coupling to a field. The results  derived via the path integral formalism for this case are also derived via perturbation theory on the corresponding time-dependent Fokker-Planck equation.

Finally we demonstrate the analytically predicted effects by numerical simulation of some simple
models where the fluctuating field has a small number of Fourier modes. 

\section{The model}

To start with we define the general class of models of interest to us in this paper.
First we consider the dynamics of a Langevin particle whose position is denoted by $\xx(t)$ diffusing in a $d$ dimensional space with a linear coupling to a fluctuating Gaussian field, taking values
of the same $d$ dimensional space,  as introduced in \cite{dean2011}. The overall energy/Hamiltonian for the system is
\begin{equation}
H= {1\over 2}\int \phi(\xx)\Delta\phi(\xx) d\xx - hK\phi(\xx(t)),
\end{equation}
the first term is the quadratic energy of a free scalar field and the second corresponds to the tracer
seeing an effective potential $h\psi =-hK\phi$. We will take $\Delta$ and $K$ to be self adjoint operators. For two operators $A(\xx,\yy)$ and $B(\xx,\yy)$ we will denote by $AB(\xx,\yy)$ their composition as operators: $(AB)(\xx,\yy)=\int A(\xx,\z)B(\z,\yy)d\z$.

For a system obeying detailed balance (and whose equilibrium state is thus given by the Gibbs-Boltzmann distribution) the particle evolves according to
\begin{equation}\label{xd}
\dot\xx(t)= -\kappa_\xx\frac{\delta H}{\delta \xx} + \sqrt{\kappa_\xx}\eeta(t)=h\kappa_\xx\nabla K\phi(\xx(t)) + \sqrt{\kappa_\xx}\eeta(t),
\end{equation}
where the noise term is Gaussian with zero mean zero and correlation function
\begin{equation}\label{xnoise}
\langle \eeta(t)\eeta(s)^T\rangle = 2T \delta(t-s)\un,
\end{equation}
where $T$ is the temperature of the system.

In the absence of a coupling between the field and the particle ($h=0$), the particle diffuses normally and the mean squared displacement at large times behaves as
\begin{equation}
\left\langle (\xx(t)-\xx(0))^2\right\rangle  \underset{t\rightarrow\infty}{\sim}  2dT\kappa_\xx t= 2d D_\xx t,\label{dodef}
\end{equation}
where $d$ is the spatial dimension and $D_\xx = T\kappa_\xx$ is the bare diffusion constant.

We take a general dissipative dynamics for the field \cite{cha2000}:
\begin{equation}\label{phid}
\dot\phi(\xx,t)=-\kappa_\phi R{\delta H\over \delta\phi(\xx )} + \sqrt{\kappa_\phi}\xi(\xx,t)=-\kappa_\phi R\Delta \phi(\xx )  
+ h \kappa_\phi RK(\xx-\xx(t)) + \sqrt{\kappa_\phi}\xi(\xx ,t),
\end{equation}
where $R$ is a self adjoint dynamical operator and $\xi$ is a Gaussian noise of zero mean which is uncorrelated in time. In order for the field to equilibrate to the Gibbs-Boltzmann distribution, the correlation function of this noise must be
\begin{equation}\label{phinoise}
\langle \xi(\xx,t)\xi(\yy,s)\rangle  =2T R(\xx-\yy)\delta(t-s).
\end{equation}

The effective diffusion constant for the particle defined via
\begin{equation}
\left\langle (\xx(t)-\xx(0))^2\right\rangle \underset{t\rightarrow\infty}{\sim} 2dT\kappa_e t= 2d D_et,
\end{equation}
 $D_e$ is the effective or {\em late time} diffusion constant.

Our model applies to many systems: a point magnetic field of magnitude $h$ diffusing in a {\em Gaussian ferromagnet} can be modelled with $\Delta(\xx,\yy) = (-\nabla_\xx^2 +m^2)\delta(\xx-\yy)$, $K(\xx,\yy)=\delta(\xx-\yy)$ and  $R(\xx,\yy)=\delta(\xx-\yy)$ for model A dynamics or $R(\xx,\yy)=-\nabla_\xx^2\delta(\xx -\yy)$ for model B conserved dynamics \cite{cha2000}. Here the field $\phi$ can correspond to the  fluctuations of a range of order parameters in a lipid membrane, in the Gaussian
approximation, where the fluctuations are weak \cite{dem2010a, dem2010b}. The order parameter
in question could be compositional fluctuations above the demixing transition in lipid bilayers or binary fluids. Here,  a delta function form of the coupling $K$ would correspond to the tracer having a 
preference for one of the lipid phases or preferential wetting for one of the phases of a binary fluid. 
As well as compositional fluctuations, $\phi$ could also
correspond to fluctuations in local lipid ordering (gel/liquid phases) and possibly orientational
order in the lipid tails. It could also correspond to  local membrane thickness in the case where the protein induces a local hydrophobic mismatch and locally alters the thickness of the bilayer. 

To model a lipid membrane where the field represents the height fluctuations and the particle is a protein coupled to membrane curvature, we may take the Helfrich Hamiltonian \cite{hel1973} $\Delta(\xx,\yy) = (\kappa\nabla_\xx^4 -\sigma\nabla_\xx^2)\delta(\xx-\yy)$, $K(\xx,\yy)=-\nabla_\xx^2\delta(\xx-\yy)$ and $R$ is given by its Fourier transform, $\tilde R(\kk)= 1/4\eta |\kk|$, where $\eta$ is the viscosity of the solvent surrounding the membrane \cite{lin2004,naji2007a}.

The computations below can be made a little more general. As mentioned in the introduction there are a number of physical cases of non-equilibrium systems where  one is not restricted to stochastic dynamics obeying detailed balance and for instance the coupling between the particle and the field may be taken non-symmetric: instead of (\ref{phid}), we will consider
\begin{equation}\label{phidbis}
\dot\phi(\xx,t) = -\kappa_\phi R\Delta \phi(\xx)  
+ \zeta h \kappa_\phi RK(\xx-\xx(t)) + \sqrt{\kappa_\phi}\xi(\xx,t),
\end{equation}
where we have introduced the parameter $\zeta$. The case of  stochastic dynamics with detailed balance is recovered for $\zeta=1$. We note that Eq. ({\ref{xd}) has been extensively studied in the case where the field $\phi$ evolves independently of the particle position, which corresponds to $\zeta=0$. This problem is referred to as the advection diffusion of a passive scalar (the particle concentration) in a fluctuating field $\phi$. It had been suggested that this form can be used to approximate the diffusion of an active tracer particle in \cite{rei2005, lei2008}, for the case of a protein weakly coupled to membrane curvature.
In this approximation  it was found that the effect of the field fluctuations could be  to increase the diffusivity of the tracer particle with respect to that obtained when it is not coupled to the fluctuating field ($h=0$). However the numerical simulations of \cite{naji2009} where the effect of the particle position on the field is taken into account showed that the diffusion is reduced with respect to the case $h=0$, in agreement with later analytical studies \cite{lei2010, dean2011}

Also as previously mentioned, a further generalization can be made to our model: since, in a non-equilibrium system, the particle and the field are not necessarily driven by the same thermal bath, they can experience different temperatures. In this case he temperatures appearing in the correlation function of the noises in Eqs. (\ref{xnoise}) and (\ref{phinoise}) will be denoted respectively $T_\xx$ and $T_\phi$.

\section{Effective diffusion equation and path integral formalism}

\subsection{Effective diffusion equation}

Our aim is to study the average value of the mean squared displacement of the particle, we thus integrate the dynamical equation  of field  Eq. (\ref{phidbis}), assuming without loss of generality that the field $\phi=0$ at time $t=0$. This gives
\begin{equation}\label{phiexpl}
\phi(\xx ,t) = \int_{-\infty}^t \ e^{-\kappa_\phi (t-s)R\Delta}
\left[\zeta h\kappa_\phi RK(\xx -\xx (s)) +\sqrt{\kappa_\phi}\xi(\xx ,s)\right]ds.
\end{equation}
Using  this result in Eq.  (\ref{xd}) we obtain the effective diffusion equation for the particle:
\begin{equation}\label{xd_expl}
\dot\xx(t) = h\kappa_\xx \nabla K\int_{-\infty}^t \ e^{-\kappa_\phi (t-s)R\Delta}
\left[\zeta h\kappa_\phi RK(\xx (t)-\xx (s)) +\sqrt{\kappa_\phi}\xi(\xx (t),s)\right]ds + \sqrt{\kappa_\xx }\eeta(t).
\end{equation}

The right hand term can be split into two parts: a deterministic part depending only on the particle trajectory, and a stochastic part which depends on the noise driving the field and on the particle trajectory:
\begin{equation}\label{gradF}
\dot\xx(t)=\sqrt{\kappa_x}\eeta(t)+\int_{-\infty}^t \FF(\xx(t)-\xx(s),t-s)ds+\XXi(\xx(t),t),
\end{equation}
where $\XXi$ is a Gaussian noise  dependent on the position in space and time, with correlation function 
\begin{equation}\label{noiseG}
\langle \XXi(\xx ,t)\XXi(\yy,s)^T\rangle = T\GG(\xx-\yy,t-s),
\end{equation}
and we have introduced the functions
\begin{eqnarray}
\FF(\xx,u)&=&\zeta h^2\kappa_\xx\kappa_\phi\nabla K e^{-\kappa_\phi u R\Delta}RK(\xx), \label{defF} \\
\GG(\xx,u)&=&- h^2\kappa_\xx ^2\nabla\nabla^T K^2e^{-\kappa_\phi|u|R\Delta}\Delta^{-1}(\xx). \label{defG}
\end{eqnarray}

We introduce the above functions for two reasons: provide more compact notation for the switch to the path integral formalism, and show explicitly which part of the following computation is general and does not depend on the expressions of $\FF$ and $\GG$. Indeed, for a different choice of functions, (\ref{gradF}) and (\ref{noiseG}) define a more general model, which may be analyzed using the method presented here.
We will need explicit expressions for $\FF$ and $\GG$; in particular their $\xx$ dependence is rather obscure. Fourier transforming allows us to write them as a sum of functions with a completely explicit $\xx$ dependence:
\begin{equation}
\FF(\xx,u)=\zeta h^2\kappa_\xx\kappa_\phi\int\frac{d^d\kk}{(2\pi)^d}i\kk e^{-\kappa_\phi u \tilde R(\kk)\tilde\Delta(\kk)}\tilde R(\kk)\tilde K^2(\kk)e^{i\kk\cdot\xx}=\int\frac{d^d\kk}{(2\pi)^d}\FF_\kk(\xx,u)
\end{equation}
and
\begin{equation}
\GG(\xx,u)=h^2\kappa_\xx^2\int\frac{d^d\kk}{(2\pi)^d}\kk\,\kk^T \frac{e^{-\kappa_\phi |u| \tilde R(\kk)\tilde\Delta(\kk)}\tilde K^2(\kk)}{\tilde \Delta(\kk)}e^{i\kk\cdot\xx}=\int\frac{d^d\kk}{(2\pi)^d}\GG_\kk(\xx,u).
\end{equation}

\subsection{Path integral formulation}

We now turn to the path-integral formalism; the first steps are analogous to those described for general stochastic dynamics in \cite{mart1973, dede1978, aron2011} and \cite{drum1982} for transport by a time dependent incompressible velocity field. The {\em partition function} for this system is:
\begin{equation}
Z=\int \prod_t\delta\left(\dot\xx(t)-\sqrt{\kappa_\xx}\eeta(t)-\int_{-\infty}^t \FF(\xx(t)-\xx(s),t-s)ds-\XXi(\xx(t),t)\right)P[\eeta]Q[\XXi][d\xx][d\eeta][d\XXi],
\end{equation}
where $P$ and $Q$ are the functional Gaussian weight for the noises. Note that we should in principle include a Jacobian in the above expression. However if we use the Ito convention the Jacobian
term is constant and independent of the path. This is because the causality of the equation of motion 
and the use of the Ito calculus mean that the transformation matrix is triangular with diagonal
terms which are constant \cite{aron2011}.    We then use an functional integral description of the $\delta$ function, introducing the  vector field $\pp$,
\begin{equation}
Z=\int \exp\left(i\int \pp(t)\cdot\left[\dot\xx(t)-\sqrt{\kappa_\xx}\eeta(t)-\int_{-\infty}^t \FF(\xx(t)-\xx(s),t-s)ds-\XXi(\xx(t),t)\right]dt\right)P[\eeta]Q[\XXi][d\xx][d\pp][d\eeta][d\XXi].
\end{equation}
Then we use the standard result that, for a Gaussian random variable of zero mean $u$, $\langle \exp(au)\rangle=\exp(a^2\langle u^2\rangle/2)$ to perform the integration over the noises to obtain
\begin{equation}
Z=\int\exp(-S[\xx,\pp])[d\xx][d\pp]
\end{equation}
where the action is given by
\begin{equation}
S[\xx,\pp]=-i\int\pp(t)\cdot\left(\dot\xx(t)-\int_{-\infty}^t \FF(\xx(t)-\xx(s),t-s)ds\right)dt+D_\xx\int|\pp(t)|^2 dt+\frac{T_\phi}{2}\int  \pp^T(t) \GG(\xx(t)-\xx(s),t-s)\pp(s)dt\, ds.
\end{equation}
This action is the sum of the action of the pure Brownian motion
\begin{equation}
S_0[\xx,\pp]=-i\int\pp(t)\cdot\dot\xx(t)dt+D_\xx\int|\pp(t)|^2 dt
\end{equation}
and the action of the interaction
\begin{equation}%\label{sint}
S_\textrm{int}[\xx,\pp]=i\int \pp(t)\cdot \FF(\xx(t)-\xx(s),t-s)\theta(t-s) dt\,ds +\frac{T_\phi}{2}\int  \pp^T(t) \GG(\xx(t)-\xx(s),t-s)\pp(s)dt\, ds.
\end{equation}
Since $\GG(-\xx,-u)=\GG(\xx,u)$, we can write this integral only with times satisfying $t\geq s$, which will be convenient for the ensuing calculations:
\begin{equation}\label{sint}
S_\textrm{int}[\xx,\pp]=i\int \pp(t)\cdot \FF(\xx(t)-\xx(s),t-s)\theta(t-s) dt\,ds +T_\phi\int  \pp^T(t) \GG(\xx(t)-\xx(s),t-s)\pp(s)\theta(t-s)dt\, ds.
\end{equation}

\subsection{Computing averages with the free action}

We will need to compute averages with the free action $S_0$. Since it is quadratic in $\xx$ and $\pp$, we just need the one point and two point correlation functions. Moreover, the position of the particle is relevant only with respect to its position at, say, $t=0$. In order to keep compact notations we define 
\begin{equation}
\xx_0(t)=\xx(t)-\xx(0).
\end{equation}
The correlation functions required are:
$\langle \xx_0(t)\rangle_0$, $\langle \pp(t)\rangle_0$, $\langle\pp(t)\pp(s)^T\rangle_0$, $\langle\xx_0(t)\pp(s)^T\rangle_0$ and $\langle\xx_0(t)\xx_0(s)^T\rangle_0$.

Using the symmetry of $S_0$, we have immediately 
\begin{eqnarray}
\langle \xx_0	(t)\rangle_0 & = & {\bf 0},\\ \langle \pp(t)\rangle_0 & = &{\bf 0}.
\end{eqnarray} 

To obtain the two point correlation functions, we use the fact that the (functional) integral of a total (functional) derivative is zero, for example
\begin{equation}
\zz=\int \frac{\delta}{\delta\xx(s)}\left(\pp(t)e^{-S_0}\right)[d\xx][d\pp]=\int\pp(t)\dot\pp(s)^Te^{-S_0}[d\xx][d\pp],
\end{equation}
dividing each side by $Z_0$, we obtain $\left\langle \pp(t)\dot\pp(s)^T\right\rangle_0 ={\bf 0}$: $\left\langle \pp(t)\pp(s)^T\right\rangle_0$ is a constant. This constant must be zero, because the action $S_0$ does not correlate $\pp$ at different times: the $\pp(t)$ are all independent. We thus have the first correlator
\begin{equation}
\langle\pp(t)\pp(s)^T\rangle_0 = {\bf 0}.
\end{equation}

We use the same technique for the other correlators:
\begin{equation}
\zz	=\int \frac{\delta}{\delta\pp(s)}\left(\pp(t)e^{-S_0}\right)[d\xx][d\pp]=\int\left(\delta(t-s){\bf 1}+\pp(t)\left[i\dot\xx(s)^T-2D_\xx\pp(s)^T\right]\right)e^{-S_0}[d\xx][d\pp],
\end{equation}
which gives, after dividing by $Z_0$ and integrating over $s\in[0,t]$,
\begin{equation}
\left\langle\xx_0(t)\pp(s)^T\right\rangle_0=i\chi_{[0,t]}(s){\bf 1},
\end{equation}
where $\chi_{A}(s)$ is the characteristic function of the set $A$ (equal to $1$ if the argument is in
$A$ and zero elsewhere). Finally the identity, 
\begin{equation}
\zz=\int \frac{\delta}{\delta\pp(s)}\left(\xx_0(t)e^{-S_0}\right)[d\xx][d\pp]=\int\xx_0(t)\left[i\dot\xx(s)^T-2D_\xx\pp(s)^T\right]e^{-S_0}[d\xx][d\pp],
\end{equation}
leads to
\begin{equation}
\left\langle \xx_0(t)\xx_0(s)^T\right\rangle_0=2 D_\xx L([0,t]\cap [0,s])\un,
\end{equation}
where $L(I)$ is the length of the interval $I\subset \mathbb{R}$, this is of course the standard result
for free Brownian motion $\left\langle \xx_0(t)\xx_0(s)^T\right\rangle_0=2 D_\xx \min(t,s)$ if $t,s\geq 0$.

\section{Perturbative calculation of the effective diffusion constant}

\subsection{Derivation of the general result}

Here we should go back to our aim of computing the effective diffusion constant $D_e$. To do this, we have to evaluate $\langle\xx_0(t_f)^2\rangle$ at a large time $t_f$. Since we do not know how to compute averages with the action $S$, we use a perturbation expansion in terms of  averages over $S_0$, which will be denoted $\langle\dots\rangle_0$:
\begin{equation}\label{average x2}
\langle\xx_0(t_f)^2\rangle=\frac{\left\langle \xx_0(t_f)^2\exp(-S_\text{int}[\xx,\pp])\right\rangle_0}{\left\langle \exp(-S_\textrm{int}[\xx,\pp])\right\rangle_0}.
\end{equation}
Averages appearing here are not easy to compute, but it is easily to compute the first nontrivial term  in expansion in the interaction action $S_\textrm{int}$. To do this, we just expand the exponential functions:
\begin{equation}\label{foe}
\langle\xx_0(t_f)^2\rangle\simeq\frac{\left\langle \xx_0(t_f)^2(1-S_\textrm{int}[\xx,\pp])\right\rangle_0}{\left\langle 1-S_\textrm{int}[\xx,\pp]\right\rangle_0}.
\end{equation}
The interaction action is linear in $\FF$ and $\GG$ and so are the averages in (\ref{foe}); these functions are the sum over Fourier modes of functions $\FF_\kk$ and $\GG_\kk$, so we can carry out the computation with only one Fourier mode, and integrate over all  modes at the  end. We note $S_{\textrm{int},\kk}$ the interaction action associated to the $\kk$-mode, and  compute $\langle S_{\textrm{int},\kk}[\xx,\pp]\rangle_0$ and $\langle\xx_0(t_f)^2S_{\textrm{int},\kk}[\xx,\pp]\rangle_0$.

In what follows we work at fixed wave-vector $\kk$, so $\tilde\Delta(\kk)$, $\tilde K(\kk)$ and $\tilde R(\kk)$ are pure numbers and to lighten the notation we will write them $\Delta$, $K$ and $R$.

Every average we have to compute is made of terms of the form $\langle \prod_{j=1}^n O_j e^{i\kk\cdot\xx}\rangle$, where $O_j$ are operators linear in $\xx$ and $\pp$. We will need the following formula, which is easy to derive from the Wick's theorem,
\begin{equation}
\left\langle \prod_{j=1}^n O_j e^{i\kk\cdot\xx}\right\rangle=e^{-\frac{1}{2}\kk^T\langle \xx\xx^T\rangle\kk}\sum_{J\subset N}i^{|J|}\left(\prod_{j\in J}\kk\cdot\langle O_j\xx\rangle\left\langle\prod_{j\notin J} O_j\right\rangle\right) ,
\end{equation} 
where $N$ is the set $\{1,\dots,n\}$ and the sum over $J$ denotes the sum over all subsets of $N$.

We start with $\langle S_{\textrm{int},\kk}[\xx,\pp]\rangle_0$. This average contains two integrated terms. The first is $\left\langle\pp(t)\cdot \FF_\kk(\xx(t)-\xx(s),t-s)\right\rangle_0$ with $t>s$, which involves 
\begin{equation}
\left\langle\pp(t)e^{i\kk\cdot(\xx(t)-\xx(s))}\right\rangle_0=i\left\langle \pp(t)(\xx(t)-\xx(s))^T\right\rangle_0\kk e^{-k^2 D_\xx(t-s)}=0,
\end{equation}
with the notation $k^2=|\kk|^2$ and we have used that $\langle \pp(t)\xx(t)^T\rangle_0=0$ because we use the Ito convention in our path integral.
For the same reason, $\left\langle\pp(t)^T \GG_\kk(\xx(t)-\xx(s),t-s)\pp(s)\right\rangle_0=0$. Hence 
\begin{equation}
\langle S_{\textrm{int},\kk}[\xx,\pp]\rangle_0=0.
\end{equation}

Now we turn to $\langle\xx_0(t_f)^2S_{\textrm{int},\kk}[\xx,\pp]\rangle_0$. We have to compute $\left\langle \xx_0(t_f)^2 \pp(t)e^{i\kk\cdot(\xx(t)-\xx(s))}\right\rangle_0$ and $\left\langle \xx_0(t_f)^2 \pp(t)\pp(s)^T e^{i\kk\cdot(\xx(t)-\xx(s))}\right\rangle_0$, with $s\leq t$. The only non-zero term in the first average is
\begin{equation}
\begin{array}{lcl}
\left\langle \xx_0(t_f)^2 \pp(t)e^{i\kk\cdot(\xx(t)-\xx(s))}\right\rangle_0 & = & 2i\langle\pp(t)\xx_0(t_f)^T\rangle_0 \langle\xx_0(t_f)(\xx(t)-\xx(s))^T\rangle_0\kk e^{-k^2 D_\xx (t-s)} \\
& = & -4D_\xx L([0,t_f]\cap [s,t])\kk e^{-k^2 D_\xx(t-s)}\chi_{[0,t_f]}(t) \\
& = & -4D_\xx (t-\max(s,0))\kk e^{-k^2 D_\xx(t-s)}\chi_{[0,t_f]}(t) .
\end{array}
\end{equation}
The second contains two non-zero terms and we get
\begin{equation}\label{averG}
\left\langle \xx_0(t_f)^2 \pp(t)\pp(s)^T e^{i\kk\cdot(\xx(t)-\xx(s))}\right\rangle_0=\left[4D_\xx (t-\max(s,0))\kk\,\kk^T-2\chi_{[0,t_f]}(s){\bf 1}\right]e^{-k^2 D_\xx(t-s)}\chi_{[0,t_f]}(t).
\end{equation}

Now we just have to integrate the above results, and since we are interested in the long time behavior, we can neglect the terms in $o(t_f)$:
\begin{equation}\label{intf}
\left\langle i\xx_0(t_f)^2\int \pp(t)\cdot \FF_\kk(\xx(t)-\xx(s),t-s)\theta(t-s)dtds\right\rangle_0=\frac{4\zeta h^2\kappa_\xx\kappa_\phi D_\xx RK^2 k^2}{(D_\xx k^2+\kappa_\phi R\Delta)^2}t_f,
\end{equation}
and 
\begin{equation}\label{intg}
\left\langle \frac{T_\phi}{2}\xx_0(t_f)^2\int  \pp^T(t) \GG_\kk(\xx(t)-\xx(s),t-s)\pp(s)dt\, ds\right\rangle_0=2T_\phi h^2\kappa_\xx^2\Delta^{-1}K^2\frac{D_\xx k^4-\kappa_\phi R\Delta k^2}{(D_\xx k^2+\kappa_\phi R\Delta)^2}t_f .
\end{equation}

Thus, gathering these two results in (\ref{foe}) gives
\begin{equation}
\langle \xx_0(t_f)^2\rangle\simeq 2d t_f\left(D_\xx-h^2\kappa_\xx k^2 K^2\frac{\kappa_\xx T_\phi D_\xx k^2+(2\zeta D_\xx-\kappa_\xx T_\phi)\kappa_\phi R\Delta}{d\Delta(D_\xx k^2+\kappa_\phi R\Delta)^2}\right).
\end{equation}
Integrating over the modes, we get for the effective diffusion constant
\begin{equation}\label{deff}
D_e=D_\xx-\frac{h^2}{d}\int\frac{d^d\kk}{(2\pi)^d}\kappa_\xx k^2\tilde K(\kk)^2\frac{\kappa_\xx T_\phi D_\xx k^2+(2\zeta D_\xx-\kappa_\xx T_\phi)\kappa_\phi\tilde R(\kk)\tilde\Delta(\kk)}{\tilde\Delta(\kk)[D_\xx k^2+\kappa_\phi \tilde R(\kk)\tilde\Delta(\kk)]^2}.
\end{equation}
This expression is our main result. The large number of parameters makes its interpretation quite difficult, so we will apply it to some special cases to show the great variety of phenomenon which could be described.

\subsection{Application to some special cases}

\subsubsection{Stochastic dynamics with detailed balance}\label{act}

We first analyze the case  where the particle and the field see the same temperature, $T_\xx=T_\phi=T$, and the dynamics obeys detailed balance, i.e. $\zeta=1$. The effective diffusion constant is thus
\begin{equation}\label{deffh}
D_e^\textrm{db}=D_\xx\left(1-\frac{h^2}{d}\int\frac{d^d\kk}{(2\pi)^d}\frac{\kappa_\xx k^2\tilde K(\kk)^2}{\tilde\Delta(\kk)[D_\xx k^2+\kappa_\phi \tilde R(\kk)\tilde\Delta(\kk)]}\right).
\end{equation}
This result is exactly what was found in \cite{dean2011} via a Kubo formula formalism which we emphasize applies only to this particular case. By inspecting Eq. (\ref{deffh}) we see that the correction to the bare diffusion constant is always negative, the diffusion is thus slowed down by its coupling to the field. The
fact that the diffusion is slowed down for all values of $h$, and not just in the regime of small $h$ can be shown explicitly within the Kubo formalism \cite{dean2011}. A physical explanation for the slowing down
of diffusion in this case can be found in studies of the drag on a particle coupled to a field. 
The reaction of the field to the particle is to create a polaron like deformation of the field about the
particle, however a moving particle has a polaron which is not symmetric with respect to the front and
rear of the particle. This deformation generates a drag force which tends to pull the particle backwards
\cite{dem2010a, dem2010b}. 

\subsubsection{Passive diffusion}

For passive diffusion, that is $\zeta=0$, and still with equal temperatures for the particle and field , we have
\begin{equation}\label{deffpass}
D_e^\textrm{pass}=D_\xx\left(1-\frac{h^2}{d}\int\frac{d^d\kk}{(2\pi)^d}\kappa_\xx k^2\tilde K(\kk)^2\frac{D_\xx k^2-\kappa_\phi\tilde R(\kk)\tilde\Delta(\kk)}{\tilde\Delta(\kk)[D_\xx k^2+\kappa_\phi \tilde R(\kk)\tilde\Delta(\kk)]^2}\right),
\end{equation}
This result shows that, depending on the speeds of evolution $\kappa_\xx$ and $\kappa_\phi$ of the particle and the field, the particle may speed up or slow down. This is in agreement with the following intuition: with a slow field we are close to diffusion in a quenched potential, which  slows down the diffusion due to trapping in local minima which are temporally persistent. However when the 
field fluctuates  quickly the field {\em fluctuations kick the particle along} thus adding to the effective 
random force the particle experiences. However this picture is not totally valid, there appears an optimal
value of $\kappa_\phi$ at which the perturbative enhancement of the tracer's diffusion constant is maximal. If the field fluctuates too quickly the effect of field fluctuations simply average out to zero. 
This result is quite difficult to understand physically but it resembles closely the 
phenomenon of stochastic resonance \cite{ben1981, hang1998, well2004}, where the application of a periodic but  deterministic potential to Brownian particles can show an optimal frequency at which the particle dispersion is maximized. Here, the optimal value
of $\kappa_\phi$  depends, in particular, on the bare diffusivity  $\kappa_\xx$ of the particle, thus two different species will react differently to the same external field. We note that this type of phenomena can be used to sort molecules \cite{alc2004}. 

\subsubsection{Particle not connected to a thermal bath}

Another case  which could have physical relevance is where the particle is not connected to a thermal bath, i.e. $T_\xx=0$ and energy enters the system only via the fluctuations of the field. Hence in the absence of coupling to the field, the particle cannot diffuse, as $D_\xx=0$. Our result shows that the fluctuations of the field will induced a non-zero diffusion constant for particle:
\begin{equation}\label{defft0}
D_e^{T_\xx=0}=T_\phi\frac{h^2}{d}\int\frac{d^d\kk}{(2\pi)^d}\frac{\kappa_\xx^2 k^2\tilde K(\kk)^2}{\kappa_\phi\tilde R(\kk)\tilde\Delta(\kk)^2}.
\end{equation}
The effective diffusion constant is now proportional the temperature seen by the field $T_\phi$: in some sense, the field acts as a thermal bath for the particle. Note the effect of field fluctuations is to speed up the diffusion from a zero diffusion constant to that given by Eq. (\ref{defft0}), this is in agreement with the 
physical intuition that the field fluctuations will help the particle to disperse. Interestingly we see that when $D_\xx=0$ the result for $D_e$ is independent of $\zeta$ and the active and passive cases have
the same diffusion constant.

\section{Numerical simulations}

In this section we test out our theoretical predictions against the numerical simulations of a toy model 
in one dimension. We should bare in mind that the simulation of the diffusion of active tracers is 
more computationally intensive than that for passive tracers. In the latter case we can simulate the diffusion of an ensemble of independent tracers on a given dynamical realization of the fluctuating field. In the former case however we must follow the diffusion of a single particle in the fluctuating field as
for a system with more than one particle the coupling to the field introduces interactions between the
particles \cite{gou1993,sac1995}.

\subsection{Numerical model}

We consider the simplest model for our numerical simulations: $d=1$ and we take a finite number of modes, with $k=\pm n$, $1\leq n\leq N$. We also take $\tilde\Delta(k)=k^2$, $\tilde K(k)=1$ and $\tilde R(k)=1$. We set $T_\phi=1$, $\kappa_\xx=1$. For this choice of parameter our result (\ref{deff}) reads
\begin{equation}\label{deffsim}
D_e=D_\xx-h^2\frac{D_\xx+\kappa_\phi(2\zeta D_\xx-1)}{\pi(D_\xx+\kappa_\phi)^2}\sum_{n=1}^N\frac{1}{n^2}.
\end{equation}

We will perform four simulations: first we consider stochastic dynamics with detailed balance, and let the coupling $h$ vary, to explore the range of validity of our perturbative result. Then, we will simulate the three special cases described above, for different values of the speed of evolution of the field $\kappa_\phi$. Each simulation is performed for 1 mode and 10 modes.

In the simulations, we let one particle evolve in the field for a long time $\tau\gg \kappa_\xx^{-1},\, \kappa_\phi^{-1}$, and we measure its position at a fixed set of times. We repeat this simulation a large number of times (around $10^5$) and, using these measurements, we compute $\langle \xx(t)^2\rangle/2dt$, where $t$ is the measurement time. For large $t$, this function fluctuates around a mean value, which gives us the effective diffusion constant. When these fluctuations are not small, they are taken into account with error-bars on the plots.

\subsection{Validity range of the perturbative result}

The first question that arises is: to what extent is our perturbative result valid? To find the validity range for $h$, we take stochastic dynamics with detailed balance, with 1 mode and $\kappa_\phi=1$ and look at the effective diffusion coefficient as a function of $h$. The comparison between the simulations and the result (\ref{deffsim}) is given Fig. \ref{validity} and it shows that our computation is valid (i.e. the relative error is less than 5 \%) for $h\lesssim 1.2$. A more relevant criterion is however to what extent the 
deviation from the bare result can be predicted by our result. Fig. \ref{validity} shows that the theory 
predicts the deviations of the diffusion constant from its bare value in the region where the diffusion
constant deviates of the order of 15-20\% from its bare value.

\begin{figure}
 \begin{center}
%\resizebox{0.8\hsize}{!}{\includegraphics[angle=0]{validity2.eps}}
\includegraphics[angle=0]{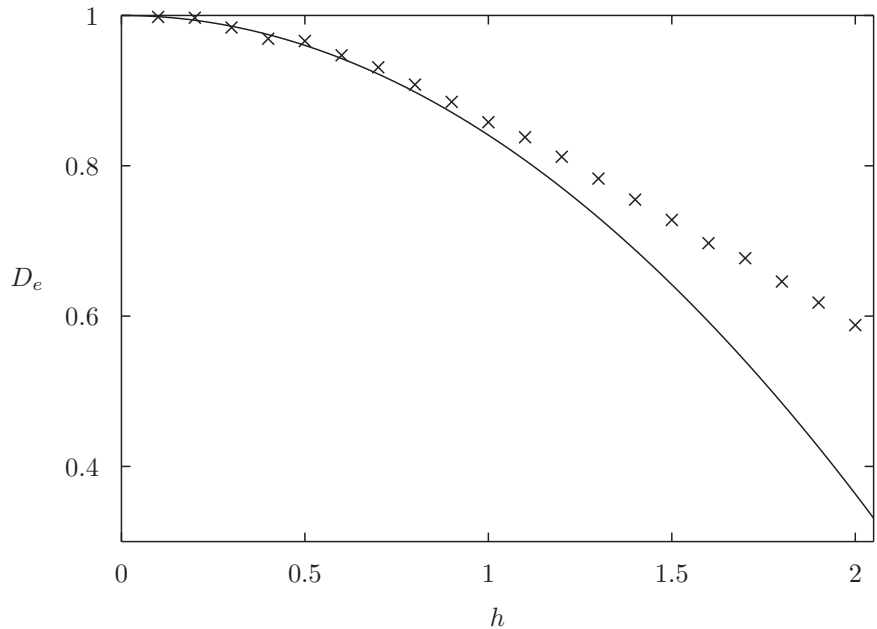}
\end{center}
\caption{Effective diffusion coefficient for stochastic dynamics with detailed balance (Gaussian
ferromagnet with model A dynamics) with a single mode field as a function of the coupling constant $h$. The crosses represent numerical simulations, the solid line is the perturbative result Eq. (\ref{deffsim}).}

%(\ref{keffhnum}) and the dashed line for the improved perturbative result (\ref{keffhimp}).}
\label{validity}
\end{figure}

\subsection{Stochastic dynamics with detailed balance}

We can also vary the rate of evolution of the field $\kappa_\phi$. We set $D_\xx=1$ and plot $D_e^\textrm{db}(\kappa_\phi)$ for 1 mode and $h=1$, and for 10 modes and $h=0.5$, comparing the numerical simulations results with (\ref{deffsim}). The results are shown in  Fig. \ref{fkeff}. For 1 mode and $h=1$, at the border of the  range of the perturbative approach, our results are in quite good agreement with the simulations.

\begin{figure}
 \begin{center}
\includegraphics[angle=0]{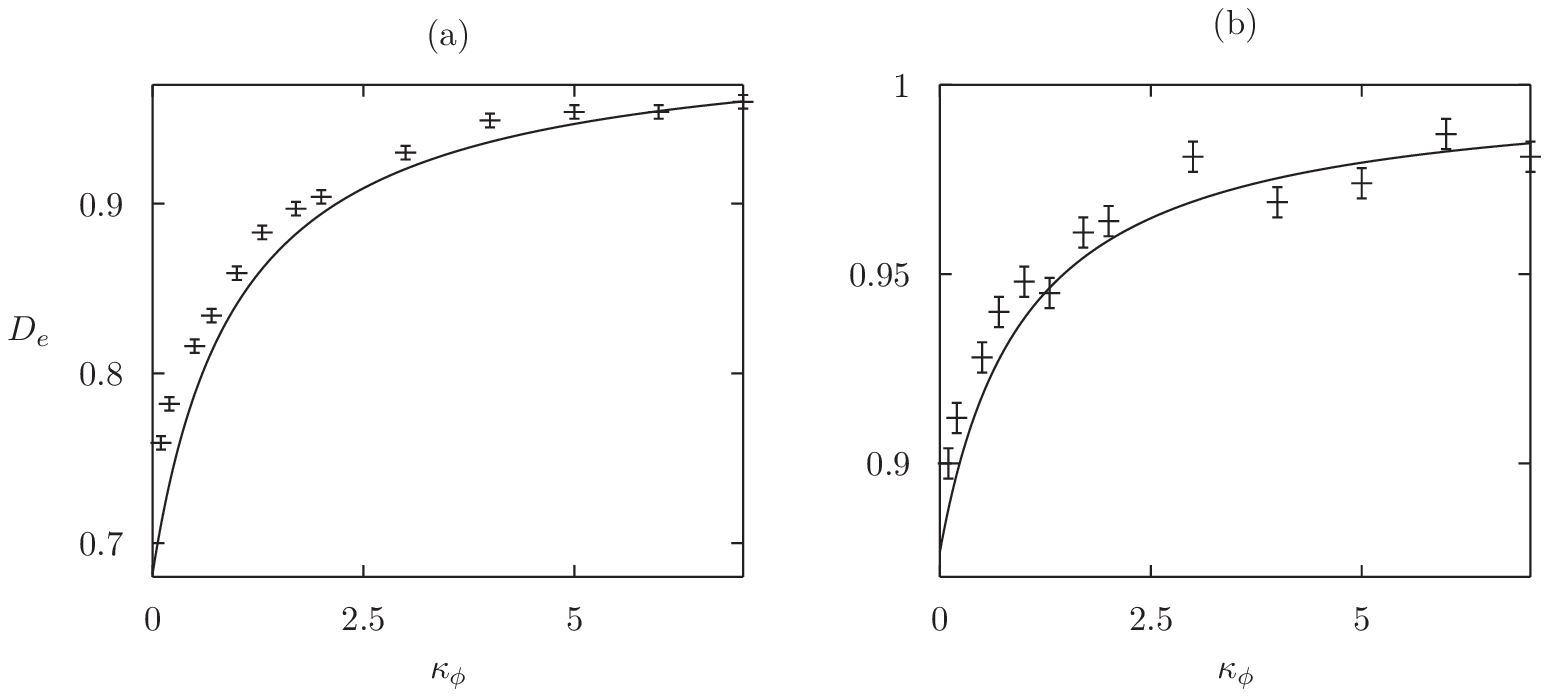}
\end{center}
\caption{Effective diffusion coefficient for stochastic dynamics with detailed balance (Gaussian
ferromagnet with model A dynamics)  as a function of $\kappa_\phi$ with $D_\xx=1$ and (a) 1 mode and $h=1$ (b) 10 modes and $h=0.5$: numerical simulations (crosses) and perturbative results (solid lines).}
\label{fkeff}
\end{figure}

\subsection{Passive diffusion}

For passive diffusion, the results of numerical simulations  are shown Fig. (\ref{fkeffpass}) and compared to Eq. (\ref{deffsim}). We see that the analytical predictions are in good agreement with the results of simulations. In particular we see that depending on the relative values of $\kappa_\xx$ and $\kappa_\phi$, the diffusion is either slowed down or speeded up. For small values of $\kappa_\phi$ the diffusion
is reduced, however on increasing $\kappa_\phi$ the diffusion speeds up and passes through a maximum before decaying towards the bare value $D_\xx=1$ as predicted by our perturbative calculations.

\begin{figure}
 \begin{center}
\resizebox{0.8\hsize}{!}{\includegraphics[angle=0]{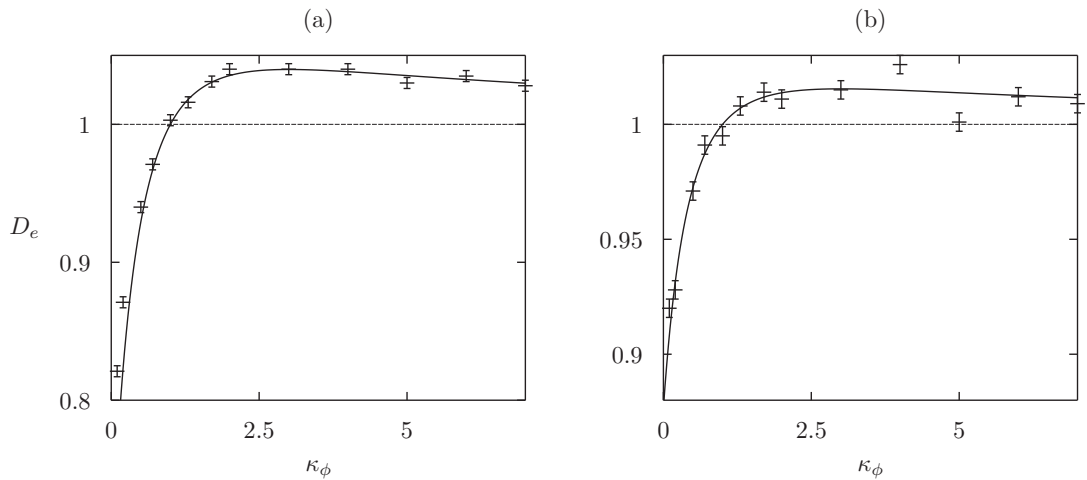}}
\end{center}
\caption{Effective diffusion coefficient for passive diffusion as a function of $\kappa_\phi$ with $D_\xx=1$ and (a) 1 mode and $h=1$ (b) 10 modes and $h=0.5$: numerical simulations (crosses) and perturbative results (solid lines).}
\label{fkeffpass}
\end{figure}

\subsection{Particle not connected to any thermal bath}

The results for numerical simulations when the particle is not connected to any thermal bath are shown Fig. \ref{fdeffT0}. They are in good agreement with Eq. (\ref{deffsim}) for $\kappa_\phi\gtrsim 1.5$. When $\kappa_\phi\rightarrow 0$, according to our computation, the effective diffusion coefficient diverges, whereas physically it should go to zero, what is confirmed by the simulations. This discrepancy comes from the fact that we neglected the terms in $o(t_f)$ in our computation of $\langle \xx(t_f)^2 S_\textrm{int}\rangle$ and we took the limit $t_f\rightarrow\infty$ before taking the limit $\kappa_\phi\rightarrow 0$.

\begin{figure}
 \begin{center}
\resizebox{0.8\hsize}{!}{\includegraphics[angle=0]{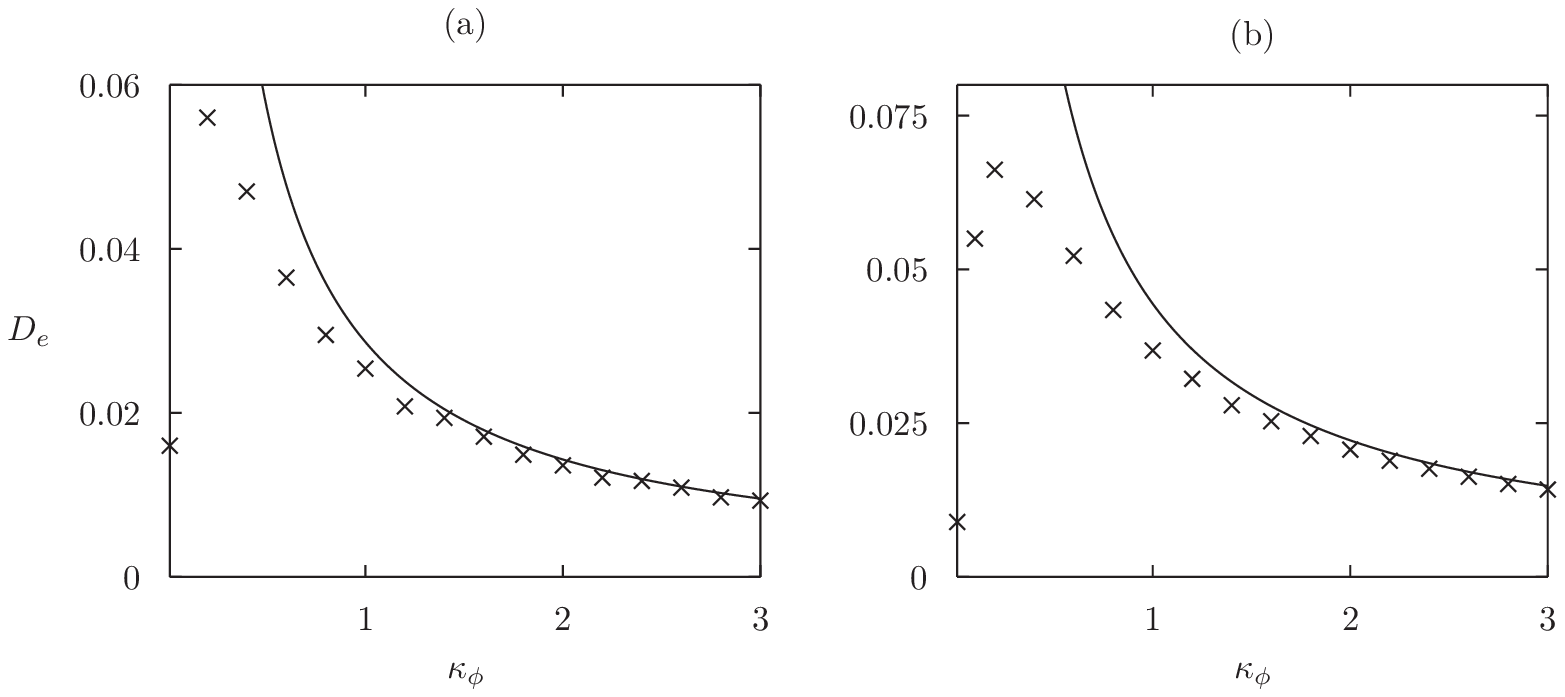}}
\end{center}
\caption{Effective diffusion coefficient for a particle not connected to any thermal bath as a function of $\kappa_\phi$ with $\kappa_\xx=1$ and (a) 1 mode and $h=0.3$ (b) 10 modes and $h=0.3$: numerical simulations (crosses) and perturbative results (solid lines).}
\label{fdeffT0}
\end{figure}

\section{Diffusion coefficient from the probability density for the passive case}

Here we use a perturbation expansion of the Fokker-Planck equation in order to compute the 
perturbative correction for passive diffusion in a fluctuating field. The basic formalism is that
the described in \cite{bou1990,dean2007}. In terms of the general model of this paper we are thus considering the  special case $\zeta=0$ and $T_\xx=T_\phi=T$.

First, we give the elementary equations for the pure Brownian motion. The probability density function $P_0(\xx,t)$ of a particle starting at $\xx=\zz$ when $t=0$, setting $P_0(\xx,t<0)=0$, satisfies
\begin{equation}
\dot P_0(\xx,t)=\nabla\cdot( D_\xx\nabla P(\xx,t))+\delta(\xx)\delta(t).
\end{equation}
We can write this equation in terms  the free diffusion operator $H_0=\partial_t-\nabla\cdot( D_\xx\nabla)$:
\begin{equation}\label{hampdf}
H_0 P_0(\xx,t)=\delta(\xx)\delta(t).
\end{equation}
and Fourier transform $\tilde P_0(\kk,\omega)=\int d\xx\, dt P_0(\xx,t)e^{-i(\kk\cdot\xx + \omega t)}$ is given by 
\begin{equation}\label{purepdf}
\tilde P_0(\kk,\omega)=\frac{1}{D_\xx k^2+i\omega}.
\end{equation}
The effective diffusion coefficient $D_e$ of a process can thus be extracted from the Fourier transform of its probability density function $\tilde P(\kk,\omega)$ via \cite{bou1990,dean2007}
\begin{equation}
D=\left(\lim_{|\kk|\rightarrow 0} k^2\tilde P(\kk,0)\right)^{-1}.
\end{equation}
We are indeed interested in the large distance behaviour of our system, that is why the effective diffusion coefficient is given by the small wave-vector behavior. With this equation in hand, our strategy is simple: compute the probability density function for the passive diffusion and extract the effective diffusion coefficient.

Now, we introduce a given field $\phi(\xx,t)$. For the general model described above, Eq. (\ref{xd}) gives the diffusion operator which replaces $H_0$ in Eq. (\ref{hampdf}):
\begin{equation}
H=H_0+H_\text{int}=\partial_t-\nabla\cdot( D_\xx\nabla+h\kappa_\xx[\nabla (K\phi)(\xx,t)]).
\end{equation}
The equation satisfied by the Fourier transform $\tilde P(\kk,\omega)$ is 
\begin{equation}
\widetilde{HP}(\kk,\omega)=\left(i\omega + D_\xx k^2\right)\tilde P(\kk,\omega)+h\kappa_\xx\int\frac{d\qq\, d\nu}{(2\pi)^{d+1}} \kk\cdot\qq\tilde K(\qq)\tilde\phi(\qq,\nu)\tilde P(\kk-\qq,\omega-\nu)=1,
\end{equation}
so $\tilde P(\kk,\omega)$ is given by the integral equation:
\begin{equation}\label{eqpdf}
\tilde P(\kk,\omega)=\tilde P_0(\kk,\omega)\left[1-h\kappa_\xx\int\frac{d\qq\, d\nu}{(2\pi)^{d+1}} \kk\cdot\qq\tilde K(\qq)\tilde\phi(\qq,\nu)\tilde P(\kk-\qq,\omega-\nu)\right].
\end{equation}
In this equation, the probability density function is that of pure Brownian motion, perturbed to the order $h$. Iterating this equation gives an explicit expression of $\tilde P(\kk,\omega)$ up to the desired order of $h$.

Once we have  the equation for $\tilde P(\kk,\omega)$ for a given field we need to extract the effective diffusion coefficient and proceed by averaging (\ref{eqpdf}) over the configurations of the field $\phi(\xx,t)$ (which does not depend on the particle position). The field has a Gaussian probability density function with a two point correlation function that can easily be computed from Eq. (\ref{phiexpl}):
\begin{equation}
\left\langle\tilde\phi(\qq,\nu)\tilde\phi(\qq',\nu')\right\rangle=\frac{2T\kappa_\phi\tilde R(\qq)}{\nu^2+(\kappa_\phi\tilde R(\qq)\tilde\Delta(\qq))^2}\times (2\pi)^{d+1}\delta(\qq+\qq')\delta(\nu+\nu').
\end{equation}

Now we have to insert this average in an explicit perturbative expansion of $\tilde P(\kk,\omega)$ given by (\ref{eqpdf}). The lowest non-zero order is the second order: the field $\phi(\xx,t)$ has to appear at least twice. Moreover, we will obtain the propability density function to the order $h^2$, what is exactly what we did with the path integral method. Computing $\tilde P(\kk,\omega)$ to the order $h^2$ and averaging the field out leads 
\begin{equation}
\left\langle\tilde P(\kk,\omega)\right\rangle=\tilde P_0(\kk,\omega)\left[1-2 h^2\tilde P_0(\kk,\omega) T\kappa_\phi\kappa_\xx^2\int\frac{d\qq\, d\nu}{(2\pi)^{d+1}}\frac{\kk\cdot\qq\, (\kk-\qq)\cdot\qq\tilde R(\qq)\tilde K(\qq)^2}{\nu^2+(\kappa_\phi\tilde R(\qq)\tilde\Delta(\qq))^2}\tilde P_0(\kk-\qq,\omega-\nu)\right].
\end{equation}

Then, we can restrict ourselves to $\omega=0$, use the expression (\ref{purepdf}) and integrate over $\nu$, using $\int\frac{d\nu}{2\pi}\frac{1}{(i\nu+\alpha)(\nu^2+\beta^2)}=\frac{1}{2\beta(\alpha+\beta)}$:
\begin{equation}
\left\langle\tilde P(\kk,0)\right\rangle=\frac{1}{D_\xx k^2}\left[1-\frac{h^2 \kappa_\xx}{k^2}\int\frac{d\qq}{(2\pi)^d}\frac{\kk\cdot\qq\, (\kk-\qq)\cdot\qq\tilde K(\qq)^2}{\tilde\Delta(\qq)(D_\xx(\kk-\qq)^2+\kappa_\phi\tilde R(\qq)\tilde\Delta(\qq))}\right].
\end{equation}

Finally, we just need to determine to behavior of the above expression when $|\kk|\rightarrow 0$; a straightforward computation gives
\begin{equation}
\left\langle\tilde P(\kk,0)\right\rangle \underset{|\kk|\rightarrow 0}{\sim} \frac{1}{D_\xx k^2}\left[1+\frac{h^2 \kappa_\xx}{d }\int\frac{d\qq}{(2\pi)^d}\frac{q^2\tilde K(\qq)^2(D_\xx q^2-\kappa_\phi\tilde R(\qq)\tilde\Delta(\qq))}{\tilde\Delta(\qq)(D_\xx q^2+\kappa_\phi\tilde R(\qq)\tilde\Delta(\qq))^2}\right],
\end{equation}
and we recover the effective diffusion coefficient given in (\ref{deffpass}):
\begin{equation}
D_e^\text{pass}=D_\xx\left(1-\frac{h^2\kappa_\xx}{d}\int\frac{d^d\kk}{(2\pi)^d} q^2\tilde K(\qq)^2\frac{D_\xx q^2-\kappa_\phi\tilde R(\qq)\tilde\Delta(\qq)}{\tilde\Delta(\qq)[D_\xx q^2+\kappa_\phi \tilde R(\qq)\tilde\Delta(\qq)]^2}\right).
\end{equation}

\section{Conclusions}

We have analyzed the diffusive behavior of a tracer particle diffusing in a time dependent Gaussian
potential in the limit of weak coupling between the particle and the field. The method has the advantage
that it can be applied to a wide range of models and not just the cases of passive diffusion or active
diffusion with detailed balance. In these two aforementioned cases the method agrees with results
obtained respectively via a perturbation expansion of the Fokker Planck equation and a Kubo formulation. We have also been able to look at non-equilibrium systems with variable feedback
of the tracer on the field and also systems where the field and the tracer are subject to thermal noise
of different temperatures. The range of behavior seen in the late time diffusion coefficient is  quite rich, and depending on the models considered, coupling to the field can either slow down or speed up the diffusion. The speeding up or slowing down of diffusion and the possibility of a form of stochastic resonance depends on the 
relative rates of the dynamics of the fluctuating field and the bare diffusion constant of the tracer.
Extensions of the work done here beyond the weak coupling approximation would be interesting to 
pursue, it is perhaps possible to apply Gaussian or mode coupling type approximations \cite{cug1996} to analyze the regime of strong interaction and perhaps even explore whether field fluctuations can 
lead to anomalous diffusion. In addition it would be interesting to see how the effects found here are 
modified when the coupling between the field and the tracer are non-linear, for instance quadratic.
Such couplings are natural in systems where the tracer does not break the symmetry of the fluctuating
field but rather enhances or suppresses its fluctuations. An example is a stiff membrane insertion which
suppresses fluctuations in membrane curvature \cite{naji2009}. A final point that would be interesting to address is what is the effect of a finite density of tracers for the active system under stochastic dynamics obeying detailed balance. As mentioned previously there will be induced interactions between the particles \cite{gou1993,sac1995} and it would be interesting to see how this modifies the effective diffusion constant.

\end{document}